\documentclass[12pt, a4paper]{article}
\usepackage[utf8]{inputenc}
\usepackage{hyperref}
\usepackage[centertags]{amsmath}
\usepackage{bm}       
\usepackage{graphicx}
\usepackage{xcolor}
\usepackage{booktabs}
\usepackage{array}
\usepackage{pdflscape}
\let\latexput\put
\usepackage{pictexwd}
\usepackage{array}
\usepackage{color}
\usepackage{multirow}
\usepackage{tablefootnote}
\usepackage{longtable}
\usepackage{verbatim}
\usepackage{authblk} 
\usepackage{tikz}
\usetikzlibrary{arrows.meta, positioning, shapes.geometric}

\newcolumntype{L}[1]{>{\raggedright\let\newline\\\arraybackslash\hspace{0pt}}p{#1}}
\newcolumntype{C}[1]{>{\centering\let\newline\\\arraybackslash\hspace{0pt}}p{#1}}
\newcolumntype{R}[1]{>{\raggedleft\let\newline\\\arraybackslash\hspace{0pt}}p{#1}}

\def\bi{\begin{itemize}}
\def\ei{\end{itemize}}

\let\put\latexput
\title{Communicating results in trials with multiple hypotheses or adaptive design features}

\author[1]{Elina Asikanius}
\author[2]{Marcel Wolbers}
\author[3]{Mouna Akacha}
\author[4]{Andreas Brandt}
\author[5,6]{Benjamin Hofner}
\author[3]{Dieter A. Häring}
\author[7, 8]{Kit C.B. Roes}
\author[9, 7]{Marc Vandemeulebroecke}
\author[10]{David Wright}
\author[4]{Joerg Zinserling}
\author[11]{Kaspar Rufibach}

\affil[1]{Finnish Medicines Agency, Turku, Finland}
\affil[2]{Data Science and Analytics,  Roche, Basel, Switzerland}
\affil[3]{Advanced Quantitative Sciences, Novartis Pharma AG, Basel, Switzerland}
\affil[4]{Biostatistics and Special Pharmacokinetics Unit, Federal Institute for Drugs and Medical Devices, Bonn, Germany}
\affil[5]{Section Data Science and Methods, Paul-Ehrlich-Institut, Langen, Germany}
\affil[6]{Department of Medical Informatics, Biometry and Epidemiology, Friedrich-Alexander-Universität Erlangen-Nürnberg, Erlangen, Germany}
\affil[7]{Science Department IQ Health, Biostatistics, Radboud University Medical Center}
\affil[8]{Medicines Evaluation Board, Netherlands}
\affil[9]{Clinical Statistics and Analytics, Bayer BCC AG, Basel, Switzerland}
\affil[10]{Respiratory and Immunology Biometrics and Statistical Innovation, BioPharmaceuticals R\&D, AstraZeneca, Cambridge, UK}
\affil[11]{Advanced Biostatistical Sciences, Biostatistics, Merck KGaA, Switzerland}

\date{}

\begin{document}

\maketitle

\newpage

\begin{abstract}
Over time, clinical trials have increasingly incorporated complex design and analysis elements such as interim analyses, adaptations, multiple endpoints, and sophisticated multiplicity schemes for multiple endpoints and/or treatment arms following the paradigm of frequentist inference. In frequentist clinical trials multiplicity can come from (at least) four sources: multiple looks at the data, multiple endpoints, multiple populations, or multiple treatment comparisons. Normally, Type 1 error control across the multiple hypotheses is implemented to control chance of false positive decisions. To achieve this advanced techniques such as adaptive designs or graphical multiple testing procedures have been developed and are used in the design of clinical trials. However, these methods focus on hypothesis testing while subsequent estimation remains crucial to allow for a benefit-risk assessment and further use of the results by various stakeholders. Through examples, we illustrate challenges in estimation and transparent communication. In general, there are no simple solutions to this conceptual and communicational challenge. The purpose of this paper is to generate awareness of these issues and initiate a discussion about how to address them moving forward.
\end{abstract}


\section*{Disclaimer}
The views expressed in this article are personal opinions of the authors and should not be understood or quoted as being representative of or reflecting the position of the regulatory agencies or organizations with which the authors are employed or affiliated.

\section{Introduction}

Over time, clinical trials have increasingly incorporated complex design and analysis elements such as interim analyses, adaptations, multiple endpoints, and sophisticated multiplicity schemes for multiple endpoints and/or treatment arms following the paradigm of frequentist inference.
From a statistical viewpoint, multiplicity can come from at least four different sources in a frequentist confirmatory clinical trial: multiple looks at the data, multiple endpoints, multiple populations, or multiple treatment comparisons. Normally, Type 1 error control across the multiple hypotheses is implemented to control chance of false positive decisions \cite{iche9, fda_adaptive_2019, fda_multiple_2022, iche20_2025}. This requirement leads to interim analyses for efficacy within group-sequential designs with the option for early stopping, multiplicity schemes for multiple endpoints including graphical multiple testing procedures, adaptive design elements, and often combinations thereof \cite{bretz2009graphical, wassmerBrannath_2025, maurer2013}. 


On the one hand, these innovations can improve the efficiency, ethical responsibility, and feasibility of clinical drug development if used responsibly. On the other hand, they pose challenges to the reporting and interpretation of results. 

Assessment of the primary and key secondary endpoints, and multiple comparisons in general are typically designed around family-wise type 1 error control in a hypothesis test. This focus on hypotheses testing is aligned with a key element of regulatory decision making: initial assessment of whether the trial confirms efficacy. However, full appreciation of clinical trial results extends well beyond this primary assessment. For regulatory approval in the EU, a positive benefit-risk (B/R) needs to be demonstrated in a methodologically reliable way. {\it Methodological reliability} here means that the results can support an interpretation with an acceptable level of uncertainty. To achieve this, simple statements on rejection or non-rejection of a number of null hypotheses under a multiple testing scheme that protects the family-wise type 1 error are not sufficient. Rather, establishing positive B/R requires a comprehensive cross-functional evaluation of the relevance of effects on both efficacy and safety. Valid estimation of the size of the treatment effect for important endpoints, as well as associated uncertainty quantification, are critical for such an evaluation. However, estimation in designs involving one or more of the above multiplicity sources, potentially combined with mid-trial adaptations and/or graphical testing procedures, may not be straightforward and prone to bias and unclear coverage properties of confidence intervals. In addition, inference for one endpoint may depend on the results for another endpoint, an aspect which is not intuitive and difficult to reflect in communication. Finally, results may be included based on their clinical relevance (e.g. overall survival in oncology), irrespective of their estimated effect size and/or whether they were type I error protected. All this taken together introduces challenges in interpretation and communication of trial outcomes. Ambiguity or inconsistency in reporting of results, within a trial and between trials and development programs, can undermine trust, compromise B/R assessment, and lead to misinterpretation of results. Namely, patients, clinicians, regulators, health technology assessment (HTA) bodies, and other stakeholders, rely on transparent and reliable information in regulatory documents on estimated effects of clinical relevance and their accuracy to make informed decisions.

The topic of communicating clinical trial results is broad, and this paper specifically discusses communication of efficacy results in regulatory documents such as the Summary of Product Characteristics (SmPC) in Europe. Hence, we focus on settings where B/R has been confirmed to be positive. The EMA's SmPC guidance \cite{smpc} provides the following criteria for information to be potentially relevant for the prescriber:
\bi
\item {\it ...statistically compelling and clinically relevant...}
\item {\it...pre-specified endpoints or clinical outcomes in major trials...}
\item {\it Such information on clinical trials should be concise, clear, relevant and balanced, and should summarise evidence from relevant trials supporting the indication.}
\ei

Although our main focus is on regulatory decision making and regulatory documents, the identified challenges are equally relevant for the communication of trial results beyond the regulatory setting. The aim of this paper is to outline common issues related to estimation in clinical trials, which we think are often not receiving the attention at the design and reporting stage that they deserve. Our ambition is to spark a discussion and initiate a change in practice through increased awareness. Such a change could also mean prioritizing a simpler over a more complicated design or multiplicity scheme, in order to be able to provide valid inference.

In Section~\ref{commchall} we outline the communication challenges that may arise from designs as described above. These challenges are then illustrated through three common examples in Section~\ref{example}. We conclude with a discussion in Section~\ref{discussion}.


\section{Communication challenges associated with multiple testing}
\label{commchall}

Decision making about an intervention integrates three aspects: 
\begin{enumerate}
    \item A decision whether a drug works, i.e. if it has a favorable effect with respect to one or more estimand(s). 
    \item Estimation of those estimands through appropriate estimators and quantification of associated uncertainties. 
    \item A holistic assessment whether the B/R profile of the drug is favorable or not. 
\end{enumerate}

The first aspect generally guides the design and definition of {\it success} of clinical trials: they are designed with a focus on hypothesis testing, typically within the Neyman-Pearson framework with pre-specification of significance level and power to reach a pre-specified treatment effect size between treatment groups for the primary efficacy endpoint. 

Over the course of developing a drug from design of a pivotal trial through HTA, the focus of stakeholders shifts from hypothesis testing to quantification of effects. Therefore, it is important that considerations of treatment effect estimation and associated uncertainty quantification as well as communication of results are not neglected at the design stage. An exclusive focus on family-wise type 1 error control is therefore insufficient. Furthermore, multiplicity control schemes often inadequately support the trial objectives and do not ensure that all required information for B/R assessment or HTA is available at the time of the respective submission, even if the schemes are valid from type 1 error control perspective. This is illustrated in Section~\ref{example3}. In the following subsections, we discuss challenges related to different types of multiple testing beyond type 1 error control.

\subsection{Per comparison versus multiplicity-adjusted uncertainty for multiple comparisons}\label{per_comparison}

A general question for the communication of uncertainty for estimates in clinical trials is whether \textit{per comparison} or \textit{multiplicity-adjusted} $p$-values and confidence intervals should be reported. Per comparison inference refers to confidence intervals and $p$-values which are valid for a single hypothesis, whereas multiplicity-adjusted inference is valid within the framework of a familywise type 1 error control scheme involving multiple comparisons. 

\textit{Per comparison inference} is typically straightforward to implement and most commonly reported. It is "weaker" than multiplicity-adjusted inference in the sense that a larger error  margin is implicitly accepted, e.g. the simultaneous non-coverage probability of $k$ per comparison $(1-\alpha)$-level confidence intervals may be as high as $k\alpha$. As a consequence, per comparison inference is not necessarily compatible with a multiplicity-adjusted hypothesis test decision. This may pose communication challenges: comparisons which are not formally significant within a multiple testing framework might still have associated per comparison $p$-values below the overall significance level and confidence intervals which exclude the null effect. 

\textit{Multiplicity-adjusted inference} is ideally compatible with the multiple testing strategy, but has several drawbacks. First, sequentially rejective multiple testing strategies rely on one-sided tests, see e.g. Bretz et al.\cite{bretz2009graphical}, whereas two-sided confidence intervals are more informative and commonly reported. This can generally be mitigated by combining multiplicity-adjusted tests in the direction of benefit and harm, respectively, as illustrated in Section~\ref{example}. 
A second, and more important, issue is that multiplicity-adjusted inference is only valid and interpretable within the framework of a pre-defined type 1 error control strategy. This implies that adjusted inference is not available for endpoints which are not part of the pre-specified multiplicity schemes. However, in some instances, it might be clinically indicated to include results for a specific endpoint in the SmPC or other communications, even though it was not part of the multiplicity scheme. For such endpoints, per comparison confidence intervals are the only option and it might be confusing to mix per comparison and (typically much wider) simultaneous confidence intervals for different endpoints in one concise document such as the SmPC. 

Another important methodological drawback is the unavailability of informative multiplicity-adjusted confidence intervals for commonly used multiple testing procedures. While informative simultaneous confidence intervals are readily available for Bonferroni-adjusted tests, hierarchical multiple testing and more complex graphical multiple testing procedures are much more common. For these latter approaches, associated multiplicity-adjusted confidence intervals are often not informative. That is, these intervals may only exclude the parameter values for which the null hypothesis is true but not any small effects, even in cases where the treatment effect is large and the multiplicity-adjusted $p$-value is small \cite{strassburger2008, schmidt2014, brannath2024}. We will illustrate this issue in Section~\ref{example}. Alternative “trade-off” approaches that reserve some type 1 error to obtain more informative simultaneous confidence intervals have been proposed in the literature \cite{schmidt2014, brannath2024}. While this is a promising idea, their operating characteristics are yet to be better understood and they are not currently used in practice.

Finally, multiplicity-adjusted inference is only interpretable within the underlying (potentially graphical) testing procedure. It appears somewhat challenging to display such graphs in a document like a drug label and expect prescribers and patients to be able to understand such inference.

Given the above considerations, the choice between per comparison or multiplicity-adjusted inference and their communication is often not straightforward. We will illustrate this further in the examples of Section~\ref{example}.

\subsection{Repeated group-sequential testing of the same hypothesis}\label{groupseq}

So far, we have only covered multiple testing of different null hypotheses, whereas in group-sequential and adaptive trials, the same null hypothesis is tested repeatedly. As a consequence, commonly reported "naive" treatment effect estimates that ignore this repeated testing may be biased, and the associated $p$-values and confidence intervals may be invalid, even if interpreted per comparison.

Regulatory guidance documents recommend the consideration of methods for adjusting estimates to reduce or remove bias associated with the potential for early stopping. They also state that in the trade-off between bias and variance of an estimator, the  expectation is generally for limited to no bias in the primary estimate \cite{iche20_2025, fda_adaptive_2019}. 
Textbooks and review articles discuss a variety of methods to adjust point estimates and uncertainty quantification for a group-sequential design \cite{wassmerBrannath_2025, robertson2023_1, robertson2023_2, robertson2025_1, robertson2025_2}. However, there is currently no consensus on the preferred method for reporting adjusted results in a group-sequential trial, and common practice is to simply report unadjusted results.

The group-sequential trial literature discusses two types of bias in treatment effect estimators. The \textit{conditional} or \textit{stage-wise bias} is the difference between the expected value of an estimator and the true treatment effect, conditional on the stage at which the trial stopped. In contrast, the \textit{unconditional} or \textit{global bias} is the difference between the expected value of an estimator averaged across all possible realizations of the trial compared to the true treatment effect. The unadjusted treatment effect estimator typically has a relatively small unconditional bias, but the conditional bias for trials stopped early for efficacy can be substantial and is in the direction of treatment effect overestimation \cite{robertson2023_2, freidlin_09}. However, Freidlin and Korn \cite{freidlin_09} demonstrated that this conditional bias is usually not much larger than the treatment effect overestimation that would be seen for an appropriate subset of similar positive fixed-sample-size trials. They conclude that for trials with a well-designed interim monitoring plan, stopping at $\geq 50\%$ information fraction has a negligible impact on estimation. 

Regulatory guidance documents do not discuss bias to this level of granularity, whereas estimators that aim to minimize the conditional or unconditional bias, respectively, might be quite different, as we show in our example in Section~\ref{example2}. We also note that conditional bias-adjusted estimators only reduce but do not remove bias, have a larger variance than unadjusted estimators, and the associated methods for confidence interval estimation are somewhat ad-hoc and may have suboptimal coverage \cite{coburger2001conditional, troendle1999}. Moreover, Marschner et al. \cite{marschner2022} showed that conditional bias-adjusted estimators may yield estimates that are not in the direction of benefit, even though the trial stopped early due to benefit. For these reasons, conditional bias-adjusted estimators are currently not seen as a good default choice for communicating trial results.

Adjusted confidence intervals and $p$-values for group-sequential trials can be based on the stage-wise ordering of the sample space of possible results. This ordering is first defined by the stage at which a trial stopped; trials that stopped at the same stage (i.e., at the same interim analysis) are then compared according to their test statistics. The sample space ordering also allows for the definition of an adjusted treatment effect estimator that fulfills the criterion of median unbiasedness (i.e., the unconditional median of the estimator is equal to the true value) \cite{wassmerBrannath_2025}. For communication purposes, reporting an adjusted estimator, confidence interval, and $p$-value that all rely on the same principles is appealing. 

The adjustment methods discussed so far apply only to group-sequential hypothesis tests for which the test decision has been reached (i.e., the trial has been "stopped"). For trials that are still ongoing, so-called repeated adjusted confidence intervals and $p$-values can be calculated to provide valid inference at any stage of the trial \cite{jennison1984, wassmerBrannath_2025}. The coverage probability of the intersection of repeated $(1-\alpha)$-level confidence intervals across multiple looks is equal to $1-\alpha$. Consequently, repeated confidence intervals are conservative. Indeed, for a group-sequential trial with local significance level $\alpha_k$ at interim analysis $k$ and overall significance level $\alpha$, the repeated $(1-\alpha)$-level confidence interval is numerically identical to a naive  $(1-\alpha_k)$-level confidence interval.
As illustrated in Section~\ref{example2}, repeated confidence intervals and $p$-values provide valid per comparison inference for a secondary endpoint for which further efficacy interim analyses are still planned at the stage where the test decision for the primary endpoint has been reached.
They can also be used if at an interim analysis for efficacy it is decided to continue the trial despite crossing the group-sequential boundary.

\subsection{More complex designs}
\label{complex_designs}

More complex trial designs, which combine graphical multiple testing approaches across endpoints and group-sequential trial methodology as described by Maurer and Bretz \cite{maurer2013}, are also commonly used. To implement such methods group-sequential boundaries for the same endpoint at different overall significance levels are typically required. In settings where the analysis time point for secondary endpoints is driven by a group-sequential boundary for the primary endpoint, care is required to in addition ensure type 1 error control for secondary endpoints \cite{hung2007, glimm2010}. Moreover, bias in estimation of effects for the primary endpoint may propagate to positively correlated secondary endpoints. While bias-adjustment methods for secondary endpoints in group-sequential trials have been published, see e.g. references in Robertson et al.\ \cite{robertson2023_1}, they are currently not widely used. More generally, it becomes increasingly challenging in such settings to disentangle the different sources of multiplicity and to communicate results appropriately. 

As a simple example that already illustrates these challenges we discuss hierarchical testing of two group-sequential endpoints in Section~\ref{example3}. 

\subsection{Terminology}

Even when restricting attention to trials with interim analyses only, terminology remains ambiguous, see Asikanius et al.\cite{asikanius_25} for proposals to standardize. More generally, terms used in the broad context of qualifying results of clinical trials are a mix of well defined statistical terminology, statistical terminology that different people understand differently, and terminology that is by definition somewhat unspecific. Examples of well defined terms are {\it type 1 error control} and {\it statistical significance}. These terms are relevant in the frequentist framework and do not have an obvious equivalent, for example, when Bayesian inference is used. This is a challenge from a regulatory perspective, because all trials should be assessed according to the same criteria and guidelines are relevant for all designs. An example of terminology that easily leads to differences in detailed meaning are words such as {\it confirmatory}, {\it descriptive}, {\it numerical result}, {\it nominal significance}, {\it valid inference}, {\it simultaneous confidence interval}, or {\it adjusted confidence interval}. All these terms are often used with good intentions in an attempt to create clarity and balanced messaging. However, for many of them there is no general consensus on their meaning, so using them likely leads to differences in interpretation depending on the stakeholder, which in turn may lead to confusion. For others, such as {\it valid inference} and {\it adjusted confidence interval}, more specific language is required. Lastly, often terminology used in regulatory guidelines is purposely not black and white. For example, whether a result is {\it methodologically reliable} or {\it statistically compelling} is context dependent and no simple criteria can be given. 

\section{Examples}
\label{example}
In this section, we use three fictional examples to illustrate communication challenges for randomized trials. The examples are common in drug development and relatively straightforward, as they use common statistical methods including group-sequential designs, hierarchical, and graphical multiple testing procedures. This illustrates that challenges already arise in the absence of complex multiple testing schemes and adaptations. For each example, we will discuss trial objectives, design, statistical methods, results, methodological and communication challenges, and recommendations to support transparent and reliable communication of trial results. For illustrative purposes, the examples focus on the communication of treatment effects and associated uncertainties whereas estimands and estimation procedures (including e.g. handling of missing data) are out of scope. 

\subsection{A randomized trial in early Alzheimer's disease with hierarchical testing}
\label{example1}

\subsubsection*{Trial objectives}
The objective of this trial is to assess the efficacy of a novel antibody in patients with early Alzheimer’s disease. The trial is designed to investigate the change from baseline to month 18 for three clinically relevant endpoints: Clinical Dementia Rating–Sum of Boxes (CDR-SB), Alzheimer’s Disease Assessment Scale–Cognitive Subscale (ADAS-Cog13), and Functional Activities Questionnaire (FAQ). These endpoints are not only important for regulatory decision making but also to inform healthcare practitioners and patients about treatment options. 

\subsubsection*{Trial design}
\begin{figure}[h!]
    \centering
    \begin{tikzpicture}[
        node distance=2cm and 3cm,
        fatnode/.style={ellipse, fill=gray!20, minimum size=10mm}]
        every edge/.style={draw, -Stealth},
        faded/.style={color=gray!50}
    ]
        \node[fatnode] (C) {CDR-SB};
        \node[fatnode, below=of  C] (A) {ADAS-Cog13};
        \node[fatnode, below=of A] (F) {FAQ};
    
        \draw [thick, ->] (C) -- (A) node[midway,left] {\Large\(    1 \)};
        \draw [thick, ->] (A) -- (F) node[midway,left] {\Large\(    1 \)};
    \end{tikzpicture}
    \caption{Hierarchical testing approach applied to the randomized trial in early Alzheimer's disease }
    \label{fig:alzheimers}
\end{figure}
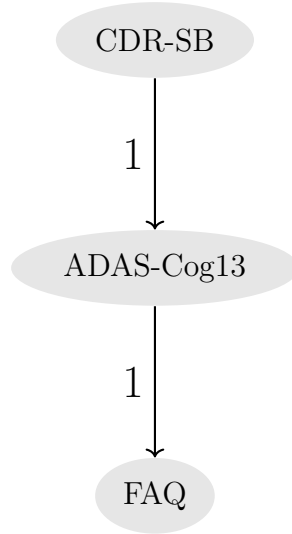

The trial is 1:1 randomized, double-blind, placebo-controlled where patients are followed up for 18 months. To control the overall one-sided type 1 error at $\alpha = 2.5\%$, a hierarchical testing procedure is implemented, whereby the endpoints are tested sequentially in the order depicted in Figure~\ref{fig:alzheimers}. 

\subsubsection*{Statistical methods}
The endpoints are analyzed using an analysis of covariance model with treatment as the main covariate and adjustment for pre-specified prognostic factors. Hypothesis testing proceeds sequentially in the order of the pre-specified hierarchy. If and only if the hypothesis test for an endpoint is significant at the one-sided 2.5\% significance level does testing continue to the next endpoint in the hierarchy. In order to obtain two-sided inference at the 5\% significance level, the hierarchical testing procedure is conducted twice, once in the direction of benefit for all endpoints and once in the direction of harm, respectively.

For each endpoint, point estimates and per comparison as well as multiplicity-adjusted two-sided simultaneous confidence intervals and $p$-values are calculated. Adjusted inference is conducted as described in Bretz et al.\ \cite{bretz2009graphical} and implemented in the R package gMCPLite \cite{gMCPLite}.

\subsubsection*{Trial results}
The results of the trial are summarized in Table~\ref{tab:alzheimers}. The primary endpoint was the change from baseline to month 18 in the CDR-SB score. The mean difference for this endpoint is $-0.40$, with a $95\%$ per comparison confidence interval ranging from $-0.70$ to $-0.10$. The unadjusted and multiplicity-adjusted $p$-values are both $0.01$, indicating a statistically significant benefit of the treatment on the primary endpoint. The multiplicity-adjusted confidence interval ranges from $-0.70$ to $0.00$, i.e. its upper bound is not informative beyond confirming the result of the hypothesis test.  

For the first secondary endpoint, the change from baseline to month 18 in ADAS-Cog13, the mean difference is $-1.30$, with a per comparison $95\%$ confidence interval from $-2.80$ to $0.20$. The unadjusted and multiplicity-adjusted $p$-values are both $0.09$. The null hypothesis of no effect cannot be rejected for this secondary endpoint. The multiplicity-adjusted confidence interval ranges from $-\infty$ to $0.20$, i.e. only its upper bound is informative. 

The third endpoint, the change from baseline to month 18 in FAQ, showed a mean difference of $-1.10$, with a per comparison $95\%$ confidence interval from $-1.90$ to $-0.30$. The unadjusted $p$-value is $0.01$ whereas the multiplicity-adjusted $p$-value is 0.09. While this endpoint appears statistically significant when considered per comparison, i.e. in isolation, the hierarchical testing strategy dictates that a hypothesis test only maintains the family-wise error rate over all three endpoints if the preceding secondary endpoint (ADAS-Cog13) is also significant. Since the ADAS-Cog13 result was not statistically significant, the FAQ result has to be considered {\it not statistically significant} according to the prespecified multiplicity scheme. The multiplicity-adjusted confidence interval for the FAQ result is $-\infty$ to $+\infty$, i.e. it is completely uninformative. 

\begin{table}[h!]
    \centering
    \caption{Treatment effect estimates (change from baseline to month 18) with \textit{per comparison} and \textit{multiplicity-adjusted} two-sided confidence intervals and $p$-values for the fictional trial in Alzheimer's disease.}
    \label{tab:alzheimers}
    \begin{tabular}{lcrr}
        \toprule
\textbf{} & \multicolumn{1}{c}{\textbf{Mean}} & \multicolumn{1}{c}{\textbf{Per comparison}} & \multicolumn{1}{c}{\textbf{Multiplicity-adjusted}} \\
        \textbf{Endpoint} & \multicolumn{1}{c}{\textbf{difference}} & \multicolumn{1}{c}{\textbf{95\% CI; $p$-value}} & \multicolumn{1}{c}{\textbf{95\% CI; $p$-value}} \\
                \midrule
        \textbf{Primary endpoint} &  & \\
        CDR-SB & $-0.40$ & [$-0.70$, $-0.10$]; $p = 0.01$ & [$-0.70$, $0.00$); $p = 0.01$ \\\hline \hline\addlinespace
        \textbf{Secondary endpoints} & & \\ \addlinespace
        ADAS-Cog13  & $-1.30$ & [$-2.80$, $+0.20$]; $p=0.09$ & [$-\infty$, $+0.20$];  $p=0.09$\\\hline \addlinespace
        FAQ & $-1.10$ & [$-1.90$, $-0.30$]; $p=0.01$ & [$-\infty$, $+\infty$]; $p=0.09$ \\
        \bottomrule
    \end{tabular}
    \begin{flushleft} \small
     Note: Two-sided multiplicity-adjusted inference was performed via two multiplicity-adjusted tests in the direction of benefit and harm, respectively, at the one-sided 2.5\% significance level. The two-sided $p$-value was calculated as twice the minimum of the two one-sided $p$-values, and the two-sided 95\% CI as the intersection of two one-sided 97.5\% CI. 
    \end{flushleft}
\end{table}


\subsubsection*{Methodological and communication challenges}
On a per comparison level, the results suggest a benefit of the intervention on the FAQ endpoint. However, this result is not statistically significant when the family-wise type 1 error should be kept, because the hierarchical testing procedure breaks at the ADAS-Cog13 endpoint higher up in the hierarchy.  This creates a communication challenge: if the FAQ results are considered for inclusion in the product label or broader communication, it is unclear how to present them without confusing stakeholders about the strength of evidence.

Multiplicity-adjusted confidence intervals can be constructed to align with the corresponding test decisions. However, they are not informative in this example and thus not suitable for communication purposes. 
As described in Section~\ref{per_comparison}, this is a well-known problem with hierarchical and graphical multiple comparison procedures. 

\subsubsection*{Communication strategy}

There are two possible communication strategies to report the results from Table~\ref{tab:alzheimers}. The first option is to report the treatment effect estimate and the per comparison confidence interval and $p$-value for the primary endpoint only. However, this omits endpoints which may be critical to informing  treatment options for healthcare practitioners. The second option is to also report estimates and per comparison confidence intervals (but not $p$-values) for the secondary endpoints. A challenge of the second approach is that reporting would need to make it explicit that the results for the secondary endpoints do not imply any confirmatory claims, i.e. that they are clearly labeled as exploratory only.   

\subsubsection*{Take-home message:} 

This example illustrates that informative simultaneous confidence intervals consistent with the multiple testing procedure are not available even in the simplest hierarchical testing settings. Careful communication is needed when the test decision and interpretation of estimates are not intuitively aligned, especially when results for secondary endpoints are considered for broader communication.

\subsection{A group-sequential trial with hierarchical testing in oncology}
\label{example2}

\subsubsection*{Trial objectives}
The objective of this trial is to assess the efficacy of a novel intervention versus the standard of care in an oncology indication. The primary endpoint is progression-free survival (PFS). Overall survival (OS) is expected to be immature for demonstration of superiority at the time of the PFS read-out. However, given their clinical relevance and the need to assess whether the treatment may adversely affect survival, the inclusion and careful interpretation of overall survival (OS) effect estimates remain essential in the submission. Subsequent analyses for OS are planned to demonstrate superiority. 

\subsubsection*{Trial design}
This is a 1:1 randomized open-label trial  with primary endpoint PFS and secondary endpoint OS. PFS is tested in a group-sequential design with an efficacy interim analysis at an information fraction of $66\%$ and a final analysis after the observation of 230 PFS events using an O’Brien-Fleming type $\alpha$-spending function at a nominal two-sided significance level of 5\%. OS is also tested at the same nominal significance level with a group-sequential design with two efficacy interim analyses, coinciding with the PFS interim and primary analysis, respectively, and corresponding information fractions of approximately $50\%$ and $75\%$, and a primary analysis after $233$ OS events. Type 1 error control at an overall two-sided significance level of $\alpha = 0.05$ is ensured through hierarchical testing
\cite{glimm2010, maurer2013}. To avoid directional errors, a significant benefit for OS can only be claimed if PFS is also significant in the direction of benefit. A significantly worse PFS effect would not allow to test OS subsequently. This is equivalent to testing one-sided hypotheses, which is necessary to control the overall type 1 error here.

%

%
%

\subsubsection*{Statistical methods}
For hypothesis testing, $p$-values from log-rank tests are compared to the local significance levels of the group-sequential design. Standard estimates and confidence intervals for the treatment effect are obtained from partial maximum likelihood estimation (MLE) of a Cox regression model with the treatment assignment as the only covariate. 

Several estimators and confidence intervals for the primary endpoint PFS which account for the group-sequential design as described in Section~\ref{commchall} are obtained for illustration purposes: The (unconditional) median unbiased estimator based on the stage-wise ordering of the sample space and corresponding confidence intervals and $p$-value are calculated with the R package rpact \cite{rpact}. 

The unconditional (global) bias-adjusted estimator introduced by Whitehead \cite{whitehead_1986} and the conditional (stage-wise) bias-adjusted estimator of Troendle and Yu  \cite{troendle1999} are calculated by implementing the simulation approach proposed by Pinheiro and deMets \cite{pinheiro1997}. These methods bias-correct the MLE. In order to obtain corresponding ad-hoc adjusted confidence intervals, the upper and lower confidence limit of the MLE is bias-corrected in the same way \cite{pinheiro1997}. 

Finally, repeated confidence intervals and $p$-values for PFS and OS are calculated with rpact \cite{rpact}. 

\subsubsection*{Trial results}

The trial results at the first efficacy interim analysis are summarized in Table~\ref{tab:oncology}. At this interim analysis, the hypothesis test for the primary endpoint PFS crossed the pre-defined efficacy boundary with an observed hazard ratio of 0.61. The median unbiased estimate and associated confidence interval for PFS are identical to the naive inference because the trial stopped at the first interim analysis. The global bias-adjusted estimator is 0.62 and also almost identical to the standard estimate. This confirms that global bias is unlikely to be an issue in this example.  The conditional bias-adjusted estimate is 0.78, i.e. somewhat attenuated towards 1, and has a substantially wider approximate confidence interval. 

OS did not cross the pre-defined efficacy boundary at the first efficacy interim analysis, but the unadjusted confidence interval from a Cox regression excludes the null effect and the naive $p$-value is below 0.05. Median unbiased estimators and bias-adjusted estimators are not available for OS because the result of the hypothesis test is not decided yet. The repeated confidence interval and the repeated $p$-value are consistent with the fact that the null hypothesis cannot (yet) be rejected. 

\begin{table}[h!]
    \centering
    \caption{\textit{Per comparison} results at the first efficacy interim analysis for the fictional oncology trial.}
    \label{tab:oncology}
    \begin{tabular}{p{7cm} > {\centering\arraybackslash}p{5cm} >{\centering\arraybackslash}p{2cm}}
        \toprule
        \textbf{Outcome / Estimator} & \textbf{Hazard Ratio (95\% CI)}& \textbf{$p$-value} \\\hline
        \midrule
        \textbf{Primary outcome: PFS} &  & \\\addlinespace
        Standard Cox regression and log-rank test & $0.61$ $[0.44,0.84]$ & $p = 0.002^a$ \\\hline \addlinespace
        Median unbiased estimator & $0.61$ $[0.44,0.84]$ & $p=0.002$ \\\hline \addlinespace
        Unconditional bias-adjusted estimator & $0.62$  $[0.44,0.84]$& $-^c$ \\\hline \addlinespace
        Conditional bias-adjusted estimator & $0.78$ $[0.45,0.98]$& $-^c$ \\\hline \addlinespace
        Repeated CI and $p$-value & $[0.41,0.92]^d$ & $p = 0.017$ \\\hline\hline\addlinespace         
        \textbf{Secondary outcome: OS} & & \\ \addlinespace
         Standard Cox regression and log-rank test & $0.67$ $[0.47,0.96]$ & $p = 0.03^b$ \\\hline \addlinespace
        Repeated CI and $p$-value &  $[0.39,1.16]^d$ & $p = 0.17$ \\     
        \bottomrule
    \end{tabular}
    \begin{flushleft} \small
        $^a$ Pre-defined efficacy boundary for PFS ($p \leq \alpha_{1,PFS}=0.012$) crossed. \\
        $^b$ Pre-defined efficacy boundary for OS ($p \leq \alpha_{1,OS}=0.006$) not crossed. \\
        $^c$ Hypothesis tests derived from bias-adjusted estimators are not commonly calculated. \\
        $^d$ Numerically identical to naive confidence intervals at level $1-\alpha_{1,PFS}$ or $1-\alpha_{1,OS}$, respectively. 
    \end{flushleft}
\end{table}

\subsubsection*{Methodological and communication challenges}

As described in Section~\ref{groupseq}, regulatory guidance documents recommend using analysis methods that adjust the primary analysis for the group-sequential design. While a wide range of estimation methods have been proposed in the literature, there is  no consensus on the preferred approach for reporting adjusted results. Consequently, regulatory bodies have not yet shifted the status quo away from simply reporting unadjusted results, although adjustment is strongly encouraged in several guidance documents for quite some time already \cite{ema_adaptive, ema_multiple_2017, fda_adaptive_2019, iche20_2025}.

Estimators that reduce the bias of the unadjusted estimator conditional on the stage at which the trial stopped are intuitively appealing. However, as noted in Section~\ref{groupseq}, they suffer from large variances and may exhibit paradoxical behavior; furthermore, rigorous confidence intervals for the corresponding estimators with guaranteed coverage are currently unavailable for them. Moreover, as pointed out by Freidlin and Korn \cite{freidlin_09}, while the conditional bias of the naive estimator in a group-sequential trial can be substantial, it is typically not much larger than the treatment effect overestimation observed in an appropriate subset of similar positive trials with a fixed sample size.

An alternative to conditional bias-adjusted estimators are those that reduce unconditional bias. Among these, the median unbiased estimator based on the stage-wise ordering of the sample space seems attractive to us as it is accompanied by a confidence interval and $p$-value derived from that same ordering.

This example covers one of the simplest settings of a group-sequential design with multiple endpoints. The communication challenges associated with hierarchical testing, discussed in Example~\ref{example1} for a fixed-sample trial, apply equally to the group-sequential setting.

We have not attempted to calculate simultaneous confidence intervals in this example. Although published methods exist to obtain simultaneous confidence intervals compatible with closed testing procedures for adaptive trials \cite{magirr2013simultaneous}, to our knowledge, they are not widely used. Furthermore, when closed tests are based on graphical multiple testing procedures, we expect them to frequently produce non-informative confidence intervals, similar to what is observed in trials with a fixed sample size.

\subsubsection*{Communication strategy}

In our example, the standard estimator, the median unbiased estimator, and the unconditional bias-adjusted estimator for the primary endpoint PFS as well as their confidence intervals are nearly identical. Therefore, it is practically irrelevant, which of them is chosen for reporting, e.g. in the SmPC. From a more principled point of view, it is preferable to report an estimator which also has a theoretical justification in the setting of a group-sequential design, e.g. the median unbiased estimator.

Given the limitations of the conditional bias-adjusted estimator mentioned in Section \ref{groupseq}, we would be hesitant to recommend it at this stage for inclusion in the SmPC. However, we feel that more debate between regulators, industry statisticians, and academics is required to reach a robust consensus on the  reporting of primary endpoints in group-sequential trials. 

If OS results are considered critical to inform  treatment options for healthcare practitioners even though the hypothesis test result for OS is not yet decided, then repeated confidence intervals appear more appropriate than unadjusted intervals. Reporting a $p$-value for OS does not seem necessary at this stage, though reporting the repeated $p$-value would not be misleading. 

\subsubsection*{Take home message}

Adjustment of estimates and confidence intervals for group-sequential testing is in principle feasible, but although strongly encouraged in regulatory guidance rarely done. A consensus on recommended methods is still lacking. Repeated confidence intervals can be used to adequately communicate results for secondary endpoints for which a confirmatory claim cannot yet be made. 

Communication issues associated with multiple testing methods across endpoints as discussed in examples~\ref{example1} and \ref{example3} persist when these methods are combined with group-sequential testing. Moreover, for trials which combine complex graphical multiple testing approaches across endpoints and group-sequential trial methodology, it is challenging to disentangle the different sources of multiplicity and to communicate results appropriately. 

\subsection{A three-arm trial of two active doses versus placebo in adults with obesity}\label{example3}

\subsubsection*{Trial objectives}
This example describes a fictional three-arm, randomized, placebo-controlled trial designed to evaluate the efficacy and safety of two doses of a new investigational treatment for obesity. The primary focus is on both short-term efficacy, assessed by percentage weight loss at week 26 (WL26), and long-term efficacy, assessed by percentage weight loss at week 52 (WL52). 
The overarching goal is to identify the dose with the most favorable B/R profile for potential inclusion in the product label. 

The timing of regulatory submission is set after the week 26 primary endpoint has been observed. Of note, as per the standard marketing authorization timelines in the EU, the results for the week 52 endpoint will become available during the assessment period. 

\subsubsection*{Trial design} The trial randomizes participants in a 1:1:1 ratio to receive either placebo, Dose 1, or Dose 2. The primary endpoint focuses on percentage weight loss at week 26, while the key secondary endpoint focuses on percentage weight loss at week 52. 

\begin{figure}[h!]
    \centering
    \begin{tikzpicture}[
        node distance=2cm and 3cm,
            fatnode/.style={ellipse, fill=gray!20, minimum size=15mm}]
            every edge/.style={draw, -Stealth},
            faded/.style={color=gray!50}
        ]
        
        \node[fatnode] (PFS1) {WL26, Dose 1, $\alpha/2$};
        \node[fatnode, right=of PFS1] (PFS2) {WL26, Dose 2, $\alpha/2$};
        \node[fatnode, below=of  PFS1] (OS1) {WL52, Dose 1};
        \node[fatnode, below=of PFS2] (OS2) {WL52, Dose 2};
        
        \draw [thick, ->] (PFS1) -- (OS1) node[midway,left] {\Large\( \frac{1}{2} \)};
        \draw [thick, ->] (PFS2) -- (OS2) node[midway,left] {\Large\( \frac{1}{2} \)};
        \draw[thick, ->] (PFS1) to[bend left=10] node[midway, above]{\Large\( \frac{1}{2} \)} (PFS2);
        \draw[thick, ->] (PFS2) to[bend left=10] node[midway, below]{\Large\( \frac{1}{2} \)} (PFS1);
        \draw[thick, ->] (OS2) to[bend left=20] node[very near start, below]{\Large\( 1 \)} (PFS1);
        \draw[thick, ->] (OS1) to[bend right=20] node[very near start, below]{\Large\( 1 \)} (PFS2);
    \end{tikzpicture}
    \caption{Graphical multiple testing scheme for the fictional obesity trial.}
    \label{fig:obesity}
\end{figure}
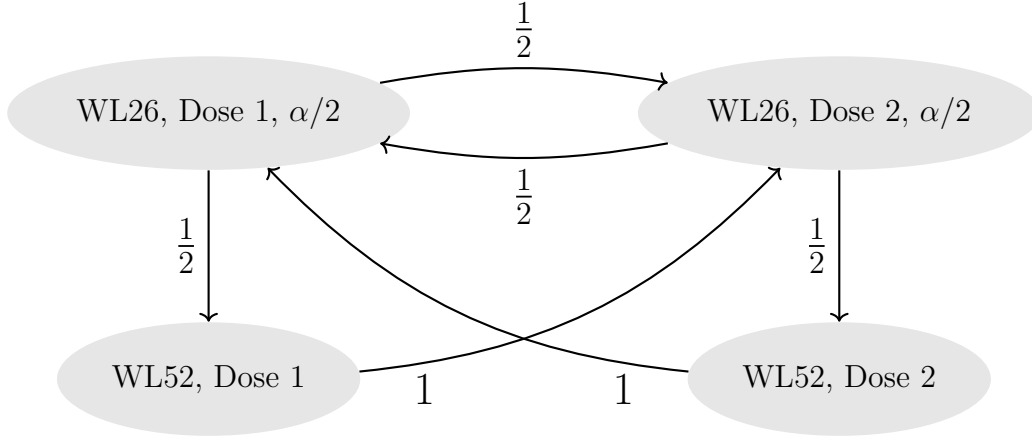

The overall one-sided significance level is set to $\alpha=2.5\%$ and to control type 1 error across multiple doses and time points, a graphical multiple testing scheme is pre-specified, see Figure~\ref{fig:obesity}. 

\subsubsection*{Statistical methods}
The endpoints are analyzed using an analysis of covariance model with treatment as the main covariate and adjustment for pre-specified prognostic factors. Hypothesis testing proceeds according to the multiple testing strategy laid out in Figure~\ref{fig:obesity}. 

For each endpoint, point estimates and per comparison as well as multiplicity-adjusted simultaneous two-sided simultaneous confidence intervals and $p$-values are calculated. Adjusted inference is conducted as described in Bretz et al.\ \cite{bretz2009graphical} and implemented in the R package gMCPLite \cite{gMCPLite}.

In order to obtain two-sided inference at the 5\% significance level, the graphical multiple comparisons procedure is conducted twice, once in the direction of benefit for all endpoints and once in the direction of harm, respectively.

\subsubsection*{Trial results}
The results of the trial are summarized in Table~\ref{tab:obesity}. Panel a. shows results at the stage where only week 26 week results are available, panel b. shows the complete data. 

At the initial analysis conducted after week 26, results from both doses are available. At this stage, accounting for the multiple testing scheme described in Figure~\ref{fig:obesity}, the primary endpoint WL26 for Dose 1 demonstrates statistical superiority over placebo at level $\alpha/2$, whereas Dose 2 is not statistically significance at level $3/4\alpha$. This can also be seen from the multiplicity-adjusted two-sided $p$-value of 0.061 for Dose 2 which is above 5\%. That is, the results available by week 26 support a confirmatory claim for Dose 1, while the evidence for Dose 2 remains less definitive. 

The situation changes when analyzing the trial at week 52. 
At this point, the secondary endpoint WL52 for Dose 1 also demonstrates statistical superiority over placebo at level $\alpha/2$, freeing the full $\alpha$ for testing Dose 2. Consequently, both the primary endpoint WL26 and the secondary endpoint WL52 reach statistical significance for both dose levels. 

\begin{table}[h!]
    \centering
    \caption{Treatment effect estimates (percentage change from baseline to week 26 or 52) with \textit{per comparison} and \textit{multiplicity-adjusted} two-sided confidence intervals and two-sided $p$-values for the fictional obesity trial. Results available by week 26 are shown in panel a, complete results in panel b.}
    \label{tab:obesity}

    \vspace{0.5em}
    \textbf{a. Partial results available by week 26}
    \begin{tabular}{lcrr}
        \toprule
        & \multicolumn{1}{c}{\textbf{Mean}} & \multicolumn{1}{c}{\textbf{Per comparison}} & \multicolumn{1}{c}{\textbf{Multiplicity-adjusted}} \\
        \textbf{Dose / Endpoint} & \multicolumn{1}{c}{\textbf{difference}} & \multicolumn{1}{c}{\textbf{95\% CI; $p$-value}} & \multicolumn{1}{c}{\textbf{95\% CI; $p$-value}} \\
                \midrule
        \textbf{Dose1} & & \\ 
        \% weight change to week 26 & $-10.0$ & $[-15.0,-5.0]$; $p < 0.001$ & $[-15.7, 0.0)$; $p<0.001$  \\ \hline\hline\addlinespace
      
        \textbf{Dose2} & & \\ 
        \% weight change to week 26 & $-5.1$ & $[-10.1,-0.1]$; $p = 0.046$ & $[-10.8, 0.2]$; $p=0.061$  \\ \addlinespace  
        \bottomrule
    \end{tabular}

    \vspace{2em}
    \textbf{b. Complete results including week 26 and 52}
    \begin{tabular}{lcrr}
        \toprule
        & \multicolumn{1}{c}{\textbf{Mean}} & \multicolumn{1}{c}{\textbf{Per comparison}} & \multicolumn{1}{c}{\textbf{Multiplicity-adjusted}} \\
        \textbf{Dose / Endpoint} & \multicolumn{1}{c}{\textbf{difference}} & \multicolumn{1}{c}{\textbf{95\% CI; $p$-value}} & \multicolumn{1}{c}{\textbf{95\% CI; $p$-value}} \\
                \midrule
        \textbf{Dose1} & & \\ 
        \% weight change to week 26 & $-10.0$ & $[-15.0,-5.0]$; $p < 0.001$ & $[-15.7, -4.3]$; $p<0.001$  \\\hline \addlinespace 
        \% weight change to week 52& $-15.0$ & $[-20.0,-10.0]$; $p < 0.001$ & $[-\infty,0.0)$; $p<0.001$ \\\hline\hline\addlinespace        
        \textbf{Dose2} & & \\ 
        \% weight change to week 26 & $-5.1$ & $[-10.1,-0.1]$; $p = 0.046$ & $[-10.8, 0.0)$; $p=0.046$  \\\hline\addlinespace  
        \% weight change to week 52& $-15.0$ & $[-20.0,-10.0]$; $p < 0.001$ & $[-\infty,0.0)$; $p=0.046$  \\
        \bottomrule
    \end{tabular}

    \begin{flushleft} \small
     Note: Two-sided multiplicity-adjusted inference was performed via two multiplicity-adjusted tests in the direction of benefit and harm, respectively, at the one-sided 2.5\% significance level. The two-sided $p$-value was calculated as twice the minimum of the two one-sided $p$-values, and the two-sided 95\% sided CI as the intersection of two one-sided 97.5\% CI. 
    \end{flushleft}

\end{table}

\subsubsection*{Methodological and communication challenges}
Challenges arise when evaluating the trial at an early time point when only week 26 results are available, because the pre-specified multiple testing scheme allows retrospective adjustment of hypothesis testing results for the primary endpoint WL26 from one dose level based on results of the secondary endpoint WL52 results from the other dose level. As a consequence, establishing benefit for the primary endpoint may be contingent on the availability of week 52 results and it may be impossible to do a proper B/R assessment and choice of the most appropriate dose based on week 26 results alone. 

Indeed, for this example, significance of the primary endpoint WL26 for Dose 2 can only be claimed after the secondary endpoint WL52 for Dose 1 is observed. In case Dose 2 had a much more favorable safety profile than Dose 1, the B/R assessment might well change between an early analysis of week 26 data only and a complete analysis of both time points. 

If the Week 52 results are considered essential for both dose selection and efficacy demonstration, aligning the timing of submission with the availability of these data is preferable. This approach ensures that the multiplicity scheme and submission strategy are consistent with the study objectives and that the most clinically relevant information is available for decision making.

As for the hierarchical testing in Example~\ref{example1}, multiplicity-adjusted confidence intervals are only partially informative and in our opinion not suitable for communication purposes. For example, the adjusted confidence intervals for the WL52 results are $[-\infty,0.0)$ for both doses and do not provide any information beyond confirming the test result. Multiplicity-adjusted $p$-values are more useful because, unlike per comparison $p$-values, they are fully compatible with the multiplicity-adjusted test results. However, displaying them jointly with per comparison confidence intervals might still be challenging. 

An additional methodological challenge is that the goal of this trial was to identify the dose with the most favorable B/R profile for potential inclusion in the product label. The observed efficacy results from Table~\ref{tab:obesity} may play an important role in the selection of the most favorable dose. This can introduce selection bias and potential overestimation for the efficacy of the selected dose. We refer to Bauer et al.\ \cite{bauer2010selection} for a more detailed discussion of this topic. 

\subsubsection*{Communication strategy}
As discussed in Example~\ref{example1}, it is recommended to clearly delineate which results are confirmatory (tested under the multiplicity scheme) and which are descriptive. 

Moving $\alpha$ from later to earlier endpoints requires strong clinical justification and is problematic from a communication perspective if it retrospectively changes the statistical significance of results which have already been reported at an earlier time point. If the week 52 result is essential for both dose selection and efficacy demonstration, the timing of submission should be aligned accordingly. Transparent discussion of the limitations of the proposed study design and submission strategy is essential. 

\subsubsection*{Take home message}
The study design, including multiplicity schemes and submission strategies, should ensure meaningful interpretation of results and be aligned with the study objectives and needs for robust decision making. Type 1 error protection is only one piece of the puzzle; communication strategies must be tailored to the clinical and regulatory context to avoid misinterpretation.

\section{Discussion}
\label{discussion}
In summary, clinical trial design based on adjusting for multiplicity in a frequentist setting puts strong emphasis on hypothesis testing and family-wise type 1 error control, while the accurate and interpretable quantification of treatment effects is often a secondary consideration. However, besides type 1 error control, an informative description of treatment effects and associated uncertainty for relevant endpoints is of interest to all stakeholders. In addition, for some trials, the focus on type 1 error control has led to multiplicity schemes which do not support the trial objectives and information generated for decision making.

One common issue is the mutual (non-)correspondence of confidence intervals and hypothesis testing. Although it holds in simple cases by definition of the confidence interval, for many only slightly more complicated testing schemes, meaningful confidence intervals that maintain this correspondence either do not even exist or require tradeoffs between significance level and coverage probability as described by Brannath et al.\cite{brannath2024}. However, current practice in clinical trials for drug approval is to use methods for hypothesis testing and confidence intervals that do not provide corresponding statements. 
Indeed, one could argue that accepting a mismatch between the decision procedure and confidence regions might be preferable to accepting non-informative intervals or sacrificing the power advantages of stepwise procedures. 
Similarly, when discussing per comparison versus multiplicity-adjusted inference, it may be preferable for a cross-functional or lay audience to report (unadjusted) effect sizes and intervals that are invariant to the testing of other endpoints -- even if this means accepting that these intervals do not have proper coverage -- rather than reporting results from multiplicity-adjusted inference, where the interpretation for a given endpoint depends on which other endpoints were measured and on the methodology applied. Of note, this does not apply when we account for multiplicity using simultaneous confidence intervals. However, these are often wide and therefore uninformative. Such a mismatch introduces communication and interpretive difficulties, particularly in settings where the coherence between the test decision and the uncertainty statement is valued. Whether and when this trade-off is worthwhile remains an open question. 
Regardless of the specific approach taken, the reporting of trial results must include honest and transparent comments and annotations, to avoid any misunderstandings. Importantly, this aspect should be considered already at the planning stage of a trial: Simpler design and analysis strategies closely linked to estimation used for decision making may be preferable over the fine tuning of the testing scheme if they reduce such issues when communicating trial results.

Admittedly, keeping apart and properly interpreting the three concepts (1) hypothesis test, (2) $p$-value, and (3) informative description of the size of the treatment effects and associated uncertainty, is not easy, even for statisticians. It is even more challenging for the various users of the clinical trial results from patients to payers who may not have a quantitative education. It is not easy to comprehend why the same numbers, a point estimate and confidence interval, have different properties depending on the context, and the same numbers may be correct in one context and incorrect in another one. We believe that focus on these aspects is of fundamental importance for multiple reasons. We do not wish to restrict the use of statistical methods in clinical trials to straightforward scenarios, on the contrary, we wish that innovation of statistical methodology continues. However, for novel methods to be accepted by regulators, they need to comprehensively address all aspects important for regulatory decision making, which goes much beyond simple hypothesis testing. Sponsors also have every interest in the correct use and interpretation of their results by all stakeholders. In addition, regulators must apply the same standards to all applicants, irrespective of which trial design, multiplicity scheme, or endpoint strategy was chosen. Hence, regulatory principles need to be generalizable. This is even more important in the current environment where we experience a shift from the pure use of frequentist statistics in clinical trials to, potentially, wider use of Bayesian methods.

 There is no easy and straightforward (frequentist) solution to the issues outlined in this paper. They are multifaceted and complex. With this paper, the authors wish to increase awareness, initiate discussion, motivate new methods development, and urge for considerations of interpretation and reporting already at the planning stage of a trial, as well as transparent documentation in protocols and statistical analysis plans.

\section{Data and Reproducibility}

No real trial data was used in this paper. All computations are reproduced using R \cite{r} in a quarto file available here: \url{https://oncoestimand.github.io/complex_designs_communication/complex_designs_communication.html}.

\section{Acknowledment}

We gratefully acknowledge many discussions and feedback on earlier versions of this manuscript from Angelika Geroldinger, Florian Klinglm\"uller, and Martin Posch.

\section{Data availability statement}

No data was used for this paper-

\section{Funding statement}

No funding was received.

\section{Conflict of interest disclosure}

None of the authors has any conflict of interest.

\section{Ethics approval statement}

No ethics approval is needed for this paper.

\bibliographystyle{abbrv}

\bibliography{biblio}

\newpage

\appendix

\end{document}